# Conductance enlargement in pico-scale electro-burnt graphene nanojunctions

Hatef Sadeghi,[1*] Jan Mol,[2] Chit Lau,[2] Andrew Briggs,[2] Jamie Warner,[2] and Colin J Lambert[1†]

[1] Quantum Technology Centre, Physics Department, Lancaster University, LA14YB Lancaster, UK and [2] Department of Materials, University of Oxford, 16 Parks Road, Oxford OX1 3PH, UK
*h.sadeghi@lancaster.ac.uk, †c.lambert@lancaster.ac.uk

**Provided the electrical properties of electro-burnt graphene junctions can be understood and controlled, they have the potential to underpin the development of a wide range of future sub-10nm electrical devices. We examine both theoretically and experimentally the electrical conductance of electro-burnt graphene junctions at the last stages of nanogap formation. We account for the appearance of a counterintuitive *increase* in electrical conductance just before the gap forms. This is a manifestation of room-temperature quantum interference and arises from a combination of the semi-metallic band structure of graphene and a crossover from electrodes with multiple-path connectivity to single-path connectivity just prior to breaking. Therefore our results suggest that conductance enlargement prior to junction rupture is a signal of the formation of electro-burnt junctions, with a pico-scale current path formed from a single $sp^2$-bond.**

Electroburning | nanoelectronics | graphene | quantum interference

## Significance

Continuation of Moore's Law to the sub-10nm scale requires the development of new technologies for creating electrode nano-gaps, in architectures which allow a third electrostatic gate. Electro-burnt graphene junctions (EGNs) have the potential to fulfil this need, provided their properties at the moment of gap formation can be understood and controlled. In contrast with mechanically-controlled break junctions, whose conductance decreases monotonically as the junction approaches rupture, we show that EGNs exhibit a surprising conductance enlargement just before breaking, which signals the formation of a pico-scale current path formed from a single $sp^2$-bond. Just as Schottky barriers are a common feature of semiconductor interfaces, conductance enlargement is a common property of EGNs and will be unavoidably encountered by all research groups working on the development of this new technology.

\body

Graphene nanojunctions are attractive as electrodes for electrical contact to single molecules [1-7], due to their excellent stability and conductivity up to high temperatures and a close match between their Fermi energy and the HOMO (highest occupied molecular orbital) or LUMO (lowest unoccupied molecular orbit) energy levels of organic materials. Graphene electrodes also facilitate electrostatic gating due to their reduced screening compared with more bulky metallic electrodes. Although different strategies for forming nano-gaps in graphene such as atomic force microscopy, nanolithography [8], electrical breakdown [9] and mechanical stress [10] have been employed, only electro-burning delivers the required gap-size control below 10 nm [11-13]. This new technology has the potential to overcome the challenges of making stable and reproducible single-molecule junctions with gating capabilities and compatibility with integrated circuit technology [14] and may provide the breakthrough that will enable molecular devices to compete with foreseeable developments in Moore's Law, at least for some niche applications [15-17].

One set of such applications is likely to be associated with room-temperature manifestations of quantum interference (QI), which are enabled by the small size of these junctions. If such interference effects could be harnessed in a single-molecule device, this would pave the way towards logic devices with energy consumption lower than the current state-of-the-art. Indirect evidence for such QI in single-molecule mechanically-controlled break junctions has been reported recently in a number of papers [18], but direct control of QI has not been possible, because electrostatic gating of such devices is difficult. Graphene electro-burnt junctions have the potential to deliver direct control of QI in single molecules, but before this can be fully achieved, it is necessary to identify and differentiate intrinsic manifestations of room temperature QI in the bare junctions, without molecules. In the present paper, we account for one such manifestation, which is a ubiquitous feature in the fabrication of pico-scale gaps for molecular devices, namely an unexpected *increase* in the conductance prior to the formation of a tunnel gap.

Only a few groups in the world have been able to implement electro-burning method to form nanogap size junctions. In a recent study of electro-burnt graphene junctions, Barreiro, et al. [19] used real-time in situ transmission electron microscopy (TEM) to investigate this conductance enlargement in the last moment of gap formation and ruled out the effects of both extra edge scattering and impurities, which reduce the current density near breaking. Also they showed that the graphene junctions can be free of contaminants prior to the formation of the nano-gap. Having eliminated these effects, they suggested that the enlargement may arise from the formation of the seamless graphene bilayers. Here we show that the conductance enlargement occurs in monolayer graphene, which rules out an explanation based on bilayers. Moreover, we have observed the enlargement in a large number of nominally identical graphene devices and therefore we can rule out the possibility of device- or flake-specific features in the electro-burning process. An alternative explanation was proposed by Lu, et al. [20], who observed the enlargement in few-layer graphene nanoconstrictions fabricated using TEM. They attributed the enlargement to an improvement in the quality of few-layer graphene due to current annealing, which simply ruled out by our experiments on electro-burnt single layers. They also attributed this to the reduction of the edge scattering due to the orientation of the edges (i.e. zigzag edges). However such edge effects have been ruled out by the TEM images of Barreiro, et al. Therefore, although this enlargement appears to be a common feature of graphene nano-junctions, so far the origin of the increase remains unexplained.

In what follows, our aim is to demonstrate that such conductance enlargement is a universal feature of electro-burnt graphene junctions and arises from quantum interference (QI) at the moment of breaking. Graphene provides an ideal platform for studying room-temperature QI effects [21], because as well as being a suitable material for contacting single molecules, it can serve as a natural two-dimensional grid of interfering pathways. By electro-burning a graphene junction to the point where only a few carbon bonds connect the left and right electrodes, one can study the effect of QI in ring- and chain-like structures that are covalently bonded to the electrodes. In this paper, we perform feedback-controlled electro-burning on single-layer graphene nano-junctions and confirm that there is an increase in conductance immediately before the formation of the tunnel-junction. Transport calculations for a variety of different atomic configurations using the non-equilibrium Green's function (NEGF) method coupled to density functional theory (DFT) show a similar behaviour. To elucidate the origin of the effect, we provide a model for the observed increase in the conductance based on the transition from multi-path connectivity to single-path connectivity, in close analogy to an optical double slit experiment. The model suggests that the conductance increase is likely to occur whenever junctions are formed from any $sp^2$-bonded material.

## Conductance through constrictions

Experimentally we study the conductance jumps by applying the method of feedback-controlled electro-burning to single-layer graphene (SLG) that was grown using chemical vapour deposition (CVD) and transferred onto a pre-patterned silicon chip (see Methods). The CVD graphene was patterned into 3 μm wide ribbons with a 200 nm wide constriction (see Fig. 1a) using electron-beam lithography and oxygen plasma etching. Feedback-controlled electro-burning has been demonstrated previously using few-layer graphene flakes that were deposited by mechanically exfoliation of kish graphite [11]. However, by applying the method to an array of nominally identical single-layer graphene devices, we can rule out the possibility of device- or flake-specific features in the electro-burning process.

We form the nano-gaps by ramping up the voltage that is applied across the graphene device. As the conductance starts to decrease due to the breakdown of the graphene, we ramp the voltage back to zero. This process is repeated until the nano-gap is formed. The $I$-$V$ traces of the voltage ramps, as shown in the figs. S1-4 of the Supplementary Information (SI), closely resemble those recorded for mechanically exfoliated graphite flakes. As the constriction narrows, the conductance of the SLG device decreases. When the conductance becomes less than the conductance quantum $G_0 = 2e^2/h$, the low-bias $I$–$V$ traces are no longer Ohmic and start exhibiting random telegraph signal (RTS) as the SLG device switches between different atomic configurations. Figure 1b shows the full $I$–$V$ trace and the final voltage ramp (inset), which exhibits a sharp increase of the conductance just before the nano-gap forms. This behaviour is characteristic of many of the devices we have studied. Out of the 279 devices that were studied, 138 devices showed a sharp increase in the conductance prior to the formation of the nano-gap ($I$–$V$ traces for 12 devices are included in the SI).

To investigate theoretically the transport characteristics of graphene junctions upon breaking, we used classical molecular-dynamics simulations to simulate a series of junctions with oxygen and hydrogen terminations as well as carbon terminated edges and then used DFT combined with non-equilibrium Green's function (NEGF) methods to compute the electrical conductance of each structure (see Methods). Figures 1c-e show three examples of the resulting junctions with oxygen terminated edges (which are the most likely to arise from the burning process), in which the left and right electrodes are connected via two (fig. 1c), one (fig. 1d) and zero (fig. 1e) pathways.

Surprisingly, the conductance $G$ through the single-path junction (fig. 1d) is larger than the conductance through the double-path junction (fig. 1c) (e.g. $G = 18\mu S$ for one path versus $G = 0.4\mu S$ for two paths in the low bias regime $V = 40mV$). For the nano-gap junction shown in fig. 1e, the conductance is less than both of these ($G = 0.016\ \mu S$). We have calculated the conductance for 42 atomic junction configurations (see figs. S6-8 of the SI), and commonly find that the conductance is larger for single-path junctions than for those with two or a few conductance paths. Approximately 40% of the total simulated junctions which were close to breaking exhibited the conductance enlargement, which is comparable with the experimental ratio of 49%.

The changes in the calculated conductances of junctions approaching rupture show a close resemblance to the experiments presented in this paper and by Barreiro, et al. [19] and arise from the changes in the atomic configuration of the junction. We therefore attribute the experimentally-observed jumps of the conductance to a transition from two- or few-path atomic configurations to single-path junctions, even though naïve application of Ohm's Law would predict a factor 2 *decrease* of the conductance upon changing from a double to a single pathway. In the remainder of this paper we will give a detailed analysis of the interference effects leading to the sudden conductance *increase* prior to the formation of a graphene nano-gap.

Before proceeding to an analysis of QI effects, we first note that the conductance enlargement cannot be attributed to changes in the band structure near breaking. The band structures of the periodic chains and ribbons shown in fig. 2 reveal that both are semi-metallic, due to the formation of a $\pi$ band associated with the $p$ orbital perpendicular to the plane of the structures. In fact, the ribbon (fig. 2b) has more open conductance channels than the chain (fig. 2a) around the Fermi energy ($E=0$). The increase in conductance upon changing from a ribbon to a chain is therefore not due to a change in band structure, but rather due to QI in the different semi-metallic pathways. A similar behaviour is also found for structures with hydrogen-termination and combined hydrogen-oxygen termination as shown in fig. S13.

Figure 3b shows the calculated current-voltage curves (corresponding transmission coefficients $T(E)$ for electrons of energy $E$ traversing the junctions are shown in fig. S10) for the five oxygen-terminated constrictions $c_1$-$c_5$ of figure 3a, with widths varying from 3 nm ($c_5$) down to a single atomic chain ($c_1$). The chains and ribbons in fig. 3a are connected to two hydrogen-terminated zigzag graphene electrodes. The blue curve of figure 3b shows that the current through the chain $c_1$ is higher than the current through the ribbon $c_2$ (green curve in fig. 3b), particularly at higher bias voltages. A non-equilibrium I-V calculation also confirms the same trend (see fig. S10b). A similar behaviour is found for structures with hydrogen-termination and without edge termination as shown in the SI (fig. S11,12). Figure 3c shows the I-V characteristic for junctions $c_1$ and $c_2$ plotted over a wider voltage range. At the penultimate stage of electroburning the $c_2$ curve is followed, until an electroburning event causes a switch from two carbon-carbon bonds to the single bond of structure $c_1$. At this point, the I-V jumps to that of structure $c_1$, as indicated in the fig. 3c by dashed line.

To demonstrate that a two-path contact between two graphene electrodes typically has a lower conductance than a single-path contact, consider a graphene nanoribbon (on the left of figs. 4a-d) connected to a carbon chain (on the right in figs. 4a and 4b) or to hexagonal chains (fig. 4c and 4d). To calculate the current flow through the junctions 4a-d and to study the effect of a bond breaking on the current when all other parameters fixed, we built a tight-binding Hamiltonian of each system (see methods). Starting from junctions 4a and 4c with two pathways between the leads, we examined the effect of breaking a single bond to yield junction 4b and 4d respectively, with only one pathway each. As shown in fig. 4, the current is increased when a bond broken (More detailed calculations are presented in the SI.) This demonstrates that in a junction formed from strong covalent bonds, the current in the one-pathway junction can be higher than in junctions with more than one pathway. This captures the feature revealed by the DFT-NEGF calculations on the structures of fig. 1, that if bonds break in a filament with many pathways connecting two electrodes from different points, the current flow can increase. This result is highly non-classical and as shown in the next section, is a consequence of constructive quantum interference in pico-scale graphene junctions connected by a single $sp^2$ bond (of length approximately 142 pm).

**Quantum interference in atomic chains and rings**

To illustrate analytically the consequences of QI in few-pathway junctions, consider the structure shown in figure 5a, which consists of an atomic chain (in fig. 5a this comprises atoms 2 and 3) connected to a single-channel lead terminating at atom $i$=1 and to a second single-channel lead terminating at atom $j$=4. Now consider adding another atomic chain in parallel to the first, to yield the structure shown in figure 5b. In physics, the optical analogue of such a structure is known as a Mach-Zehnder interferometer [22].

In the following, we shall show that the single-path structure of fig. 5a has the highest of the three conductances. This trend is the opposite of what would be expected if the lines were classical resistors (see SI), and the circles were perfect connections. In that case (a) would have the lowest conductance and (c) the highest conductance. An intuitive understanding of why our case is different begins by noting that in the quantum case, electrical conductance is proportional to the transmission coefficient $T(E)$ of de Broglie waves of energy $E$ passing through a given structure. If we neglect the lattice nature of the system, and consider the paths simply as classical waveguides, then for a wave propagating from the left hand end in each case, the bifurcations in (b) and (c) present an impedance mismatch, so that a fraction of the wave is reflected. Considering a waveguide of impedance $Z$ with a bifurcation into two waveguides, for unit incident amplitude the total transmitted amplitude is ($2\sqrt{2}/3$), and the transmitted intensity is $T = 8/9$. A similar analysis can be applied to a 1-D lattice formed of $M$ semi-infinite chains. This is illustrated in fig. 6a for $M = 2$ (a continuous chain) and fig. 6b for $M = 3$ (a bifurcation).

Within a tight-binding or Hückel description of such systems, the transmission and reflection amplitudes $r$ and $t$ are obtained from matching conditions at site "0". Then for electron energies $E$ at the band centre (ie HOMO-LUMO gap centre, which coincides with the charge neutrality point in our model), it can be shown (see SI) that the transmission coefficient $T=|t|^2$ is given by

$$T = \frac{4(M-1)}{M^2} \qquad (1)$$

For $M = 2$, this formula yields $T=1$, as expected, because system 6a is just a continuous chain with no scattering. Since $T$ cannot exceed unity, any changes can only serve to decrease $T$. For a bifurcation ($M = 3$), equation (1) yields $T = 8/9$, which is the same result as a continuum bifurcated waveguide.

When the two branches of fig. 6b come together again to form a ring, there can be further interference effects, associated with additional reflections where the branches rejoin. These may serve to decrease or increase the transmission. At most the transmission will increase to $T = 1$, but in general $T$ will remain less than unity. It might be expected that the asymmetrical ring in fig. 5c will be more likely to manifest destructive interference than the symmetrical ring in fig. 5b. These intuitive conclusions from continuous and discrete models are confirmed by the following rigorous analysis based on a tight-binding model of the actual atomic configurations, which captures the key features of the full DFT-NEGF calculations.

We consider a simple tight-binding (Hückel) description, with a single orbital per atom of 'site energy' $\varepsilon_0$ and nearest neighbour couplings $-\gamma$. As an example, for an infinite chain of such atoms, the Schrodinger's equation takes the form: $\varepsilon_0 \varphi_j - \gamma \varphi_{j-1} - \gamma \varphi_{j+1} = E\varphi_j$ for $-\infty < j < \infty$. The solution to this equation is $\varphi_j = e^{ikj}$, where $-\pi < k < \pi$ is wave vector. Substituting this into the Schrodinger's equation yields the dispersion relation of $E = \varepsilon_0 - 2\gamma \cos k$. This means that such a 1d chain possesses a continuous band of energies between $E^- = \varepsilon_0 - 2\gamma$ and $E^+ = \varepsilon_0 + 2\gamma$. Since the 1-d leads in fig. 5 are infinitely long and connected to macroscopic reservoirs, systems 5a-c are open systems. In these cases, the transmission coefficient $T(E)$ for electrons of energy $E$ incident from the first lead is obtained by noting that the wave vector $k(E)$ of an electron of energy $E$ traversing the ring is given by $k(E) = cos^{-1}(\varepsilon_0 - E)/2\gamma$. When $E$ coincides with the mid-point of the HOMO-LUMO gap of the bridge, ie when $E = \varepsilon_0$, this yields $k(E) = \pi/2$. Since $T(E)$ is proportional to $|1 + e^{ikL}|^2$, where $L$ is the difference in path lengths between the upper and lower branches, for structure 5b, one obtains constructive interference, because $e^{ikL} = e^{i0} = 1$ and for structure 5c destructive interference, because $e^{ikL} = e^{i2k} = -1$. This result is unsurprising, because it is well known that the meta-coupled ring 5c should have a lower conductance than the para-coupled ring 5b [23]. More surprising is the fact the single-chain structure 5a has a higher conductance than both 5b and 5c. To illustrate this feature, we note (see SI for more details) that the ratio of the Green's function $G_{ring}$ of the structure of fig. 5b to the Green's function of the chain 5a, evaluated between the atoms 1 and 4 is:

$$\frac{G_{ring}}{G_{chain}} = \frac{1}{2}[1 - \alpha] \quad (2)$$

where α is a small self-energy correction due to the attachment of the leads. For small α, this means that the transmission of the linear chain at the gap centre is *4x* higher than the transmission of a para ring (because transmission is proportional to the square of the Green's function), which demonstrates that the conductances of both the two-path para and meta coupled structures are lower than that of a single-path chain. This result is the opposite of the behaviour discussed in [24], where the conductance of two identical parallel chains was found to be *4x higher* than that of a single chain. The prediction in ref. [24] is only applicable in the limit that the coupling of the branches to the nodes is weak, whereas in sp$^2$-bonded graphene junctions, the coupling is strong.

## Conclusion

We have addresses a hitherto mysterious feature of electro-burnt graphene junctions, namely a ubiquitous conductance enlargement at the final stages prior to nanogap formation. Through a combined experimental and theoretical investigation of electro-burnt graphene nanojunctions, we have demonstrated that conductance enlargement at the point of breaking a consequence of a transition from multiple-path to single-path quantum transport. This fundamental role of quantum interference was demonstrated using calculations based on DFT-NEGF methods, tight-binding modelling and analytic results for the structures of fig. 5. Therefore our results suggest that conductance jumps provide a tool for characterising the atomic-scale properties of sp$^2$-bonded junctions and in particular, conductance enlargement prior to junction rupture is a signal of the formation of electro-burnt junctions, with a current path formed from a single sp$^2$-bond. Conductance enlargement is common, but does not occur in all electro-burnt nanojunctions, because direct jumps from two-path to broken junctions can occur. With greater control of the electro-burning feedback, our analysis suggests that one could create carbon-based atomic chains and filaments, which possess many of the characteristics of single molecules without the need for anchor groups, because the chains are already covalently bonded to electrodes.

## Computational Methods

The Hamiltonian of the structures described in this paper were obtained using density functional theory as described below or constructed from a simple tight-binding model with a single orbital per atom of site energy $\varepsilon_0 = 0$ and nearest neighbour couplings $\gamma = -1$.

**DFT calculation**: The optimized geometry and ground state Hamiltonian and overlap matrix elements of each structure was self-consistently obtained using the SIESTA [25] implementation of density functional theory (DFT). SIESTA employs norm-conserving pseudo-potentials to account for the core electrons and linear combinations of atomic orbitals to construct the valence states. The generalized gradient approximation (GGA) of the exchange and correlation functional is used with the Perdew-Burke-Ernzerhof parameterization (PBE) [26] a double-ζ polarized (DZP) basis set, a real-space grid defined with an equivalent energy cut-off of 250 Ry. The geometry optimization for each structure is performed to the forces smaller than 40 meV/Ang. For the band structure calculation, given structure was sampled by a 1x1x500 Monkhorst-Pack *k*-point grid.

**Transport calculation**: The mean-field Hamiltonian obtained from the converged DFT calculation or a simple tight-binding Hamiltonian was combined with our implementation of the non-equilibrium Green's function method (the GOLLUM [27]) to calculate the phase-coherent, elastic scattering properties of the each system consist of left (source) and right (drain) leads and the scattering region. The transmission coefficient *T(E)* for electrons of energy *E* (passing from the source to the drain) is calculated via the relation:

$$T(E) = Trace(\Gamma_R(E) G^R(E) \Gamma_L(E) G^{R\dagger}(E)) \quad (3)$$

In this expression, $\Gamma_{L,R}(E) = i\left(\sum_{L,R}(E) - \sum_{L,R}^{\dagger}(E)\right)$ describe the level broadening due to the coupling between left (L) and right (R) electrodes and the central scattering region, $\sum_{L,R}(E)$ are the retarded self-energies associated with this coupling and $G^R = (ES - H - \sum_L - \sum_R)^{-1}$ is the retarded Green's function, where *H* is the Hamiltonian and *S* is overlap matrix. Using obtained transmission coefficient ($T(E)$), the conductance could be calculated by Landauer formula $(G = G_0 \int dE\, T(E)(-\partial f/\partial E))$ where $G_0 = 2e^2/h$ is the conductance quantum. In addition, the current through the device at voltage *V* could be calculated as:

$$I(V) = \frac{2e}{h} \int_{-\frac{V}{2}}^{+\frac{V}{2}} dE\, T(E)[f\left(E - \frac{V}{2}\right) - f\left(E + \frac{V}{2}\right)] \quad (4)$$

where $f(E) = (1 + \exp((E - E_F)/k_B T))^{-1}$ is the Fermi-Dirac distribution function, *T* is the temperature and $k_B$= 8.6x10$^{-5}$ eV/K is Boltzmann's constant.

**Molecular dynamics**: Left and right leads (figs. 1c-e) were pulled in the transport direction by -0.1Å and 0.1Å every 40fs (200 time steps) using the molecular dynamic code LAMMPS [28]. Energy minimization of the system was achieved in each 200 time steps by iteratively adjusting atomic coordinates using following parameters: the stopping energy of 0.2, the force tolerances of 10$^{-8}$, the maximum minimizer iterations of 1000 and the number of force/energy evaluations of 10000. The atoms were treated in the REAX force field model with reax/c parameterization and charge equilibration method as described in [28] with low and high cut-off of 0 and 10 for Taper radius and the charges equilibrated precision of 10$^{-6}$. The atomic positions are updated in 0.02fs time steps at 400K with constant volume and energy. The snapshot of the atomic coordinates was sampled every 665 time steps. The whole procedure performed twice and totally 42 configuration extracted. Each of obtained set of coordinates was used as an initial set of coordinates for the subsequent self-consistent DFT loops as described above.

## Experimental Methods

Similar to previous studies using few-layer graphene flakes, the feed-back controlled electro-burning is performed in air at room temperature. The feedback-controlled electro-burning of the SLG devices [29] is based on the same method as previously used for electro-burning of few-layer graphene flakes [11] and electro-migration of metal nanowires [30]. A voltage (V) applied between the two metal electrodes is ramped up at a rate of 0.75 V/s, while the current (I) is recorded with a 200 µs sampling rate. When the feedback condition, which is set at a drop ΔI of the current within the past 15 mV is met the voltage is ramped down to zero at a rate of 225 V/s. After each voltage ramp the resistance of the SGL device is measured and the process is repeated until the low-bias resistance exceeds 500 MΩ. To prevent the SGL device from burning too abruptly at the initial voltage ramps we adjust the feedback condition for the each voltage ramp depending on the voltage at which the previous current drop occurred. The feedback conditions used were $\Delta I_{set}$ = 6, 9, 12 and 15 mA for $V_{th}$ = 1.9, 1.6, 1.3 and 1.0 V respectively.

**Acknowledgment.** This work is supported by the UK EPSRC, EP/K001507/1, EP/J014753/1, EP/H035818/1, EP/J015067/1 and from the EU ITN MOLESCO 606728, A*STAR, Oxford Martin School, and Templeton World Charity Foundation.

**Figures caption:**

**Fig. 1.** (a) Scanning electron micrograph of the graphene device, (b) Measured current-voltage characteristic of the full I–V trace. Inset: the I-V trace of the final voltage ramp prior to the formation of the nano-gap. This exhibits a sharp increase of the conductance just before the nano-gap forms. (c-e) three atomic configurations with two (c), one (d) and zero (e) pathways.

**Fig. 2.** Band structure of (a) C-O atomic chain, (b) C-O benzene chain. Grey atoms are carbon; red atoms are oxygen.

**Fig. 3.** (a) Ideal configuration with reduced junction width down to the atomic chain, (b) Calculated current-voltage relations in oxygen-terminated junctions, (c) the I-V characteristic for junctions $c_1$ and $c_2$ over a wider voltage range. Dashed lines and arrows indicate the current jump from double bond of structure $c_2$ to that of structure $c_1$ when an electroburning event occurs.

**Fig. 4.** Each of figs a-d show an electrode formed from a graphene nanoribbon (on the left) in contact with an electrode (on the right) formed from a linear chain (a and b) or a chain of hexagons (c and d). For (a) and (c) the contact to the chain is via a two bond. For (b) and (d) the contact to the chain is via single bonds. For a voltage v=20mV, the red circles show the current through each structure. The arrows indicate that upon switching from a two-bond contact to a single-bond contact, the current increases. $I_0$=77.4 μA is the current carried by a quantum of conductance $G_0$ at 1 volt.

**Fig. 5.** (a) a 1-d chain connected to 1d semi-infinite leads on the left and right, (b) two parallel chains forming a ring with para coupling to the leads and (c) two parallel chains with meta coupling to the leads

**Fig. 6.** (a) a system with M=2 semi-infinite chains, centered on site 0. (b) A system with M=3 semi-infinite chains, centered on site 0. In each case, a plane wave from the left is either reflected with reflection amplitude *r*, or transmitted with transmission amplitude

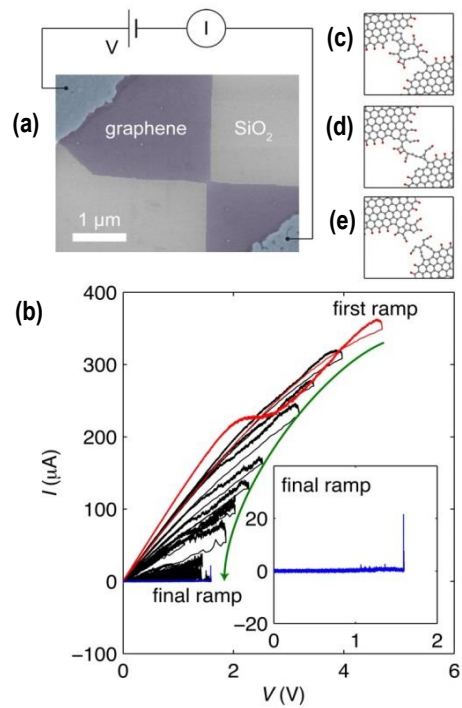

**Fig. 7.** (a) Scanning electron micrograph of the graphene device, (b) Measured current-voltage characteristic of the full I–V trace. Inset: the I-V trace of the final voltage ramp prior to the formation of the nano-gap. This exhibits a sharp increase of the conductance just before the nano-gap forms. (c-e) three atomic configurations with two (c), one (d) and zero (e) pathways.

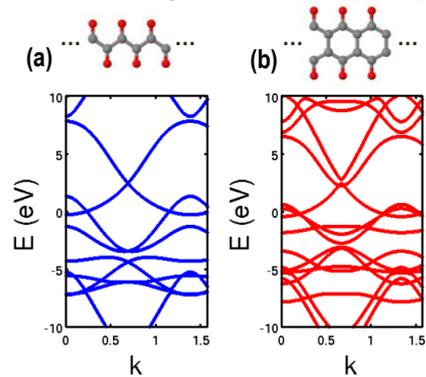

**Fig. 8.** Band structure of (a) C-O atomic chain, (b) C-O benzene chain. Grey atoms are carbon; red atoms are oxygen.

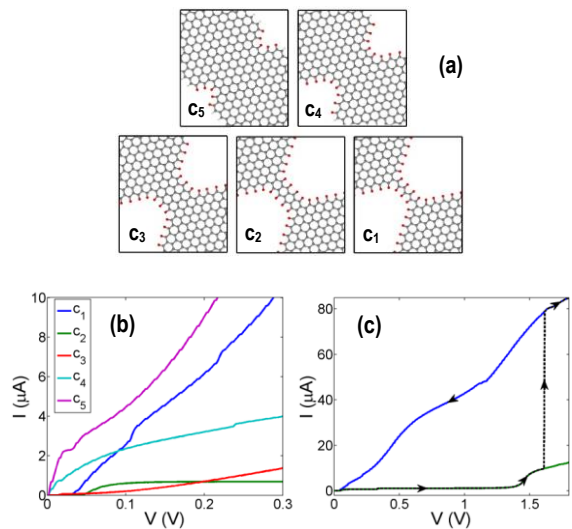

**Fig. 9.** (a) Ideal configuration with reduced junction width down to the atomic chain, (b) Calculated current-voltage relations in oxygen-terminated junctions, (c) the I-V characteristic for junctions $c_1$ and $c_2$ over a wider voltage range. Dashed lines and arrows indicate the current jump from double bond of structure $c_2$ to that of structure $c_1$ when an electroburning event occurs.

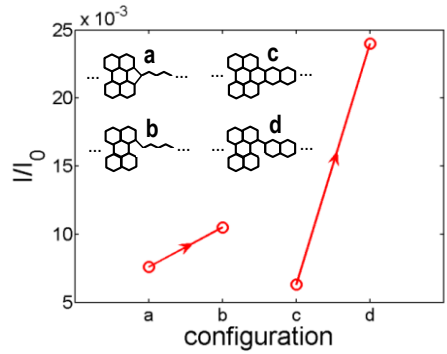

**Fig. 10.** Each of figs a-d show an electrode formed from a graphene nanoribbon (on the left) in contact with an electrode (on the right) formed from a linear chain (a and b) or a chain of hexagons (c and d). For (a) and (c) the contact to the chain is via a two bond. For (b) and (d) the contact to the chain is via single bonds. For a voltage v=20mV, the red circles show the current through each structure. The arrows indicate that upon switching from a two-bond contact to a single-bond contact, the current increases. $I_0$=77.4 µA is the current carried by a quantum of conductance $G_0$ at 1 volt.

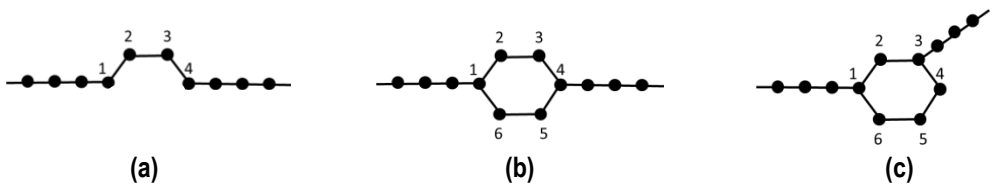

**Fig. 11.** (a) a 1-d chain connected to 1d semi-infinite leads on the left and right, (b) two parallel chains forming a ring with para coupling to the leads and (c) two parallel chains with meta coupling to the leads

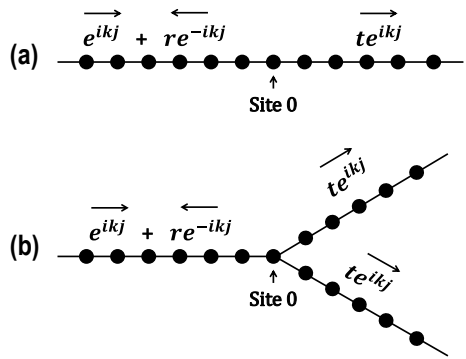

**Fig. 12.** (a) a system with M=2 semi-infinite chains, centered on site 0. (b) A system with M=3 semi-infinite chains, centered on site 0. In each case, a plane wave from the left is either reflected with reflection amplitude *r*, or transmitted with transmission amplitude